\documentclass[prb,twocolumn,aps,showpacs]{revtex4}
\usepackage{graphics,graphicx,epsfig}

\begin{document}
\title{Crossover in the scaling of island size and capture zone distributions}
\author{T. J. Oliveira${}^{1,(a)}$ and F. D. A. Aar\~ao Reis${}^{2,(b)}$
\footnote{a) Email address: tiago@ufv.br\\
b) Email address: reis@if.uff.br}}
\affiliation{${}^{1}$ Departamento de F\'isica, Universidade Federal de Vi\c cosa, 36570-000, Vi\c cosa, MG, Brazil\\
${}^{2}$ Instituto de F\'\i sica, Universidade Federal Fluminense, Avenida Litor\^anea s/n, 24210-340 Niter\'oi RJ, Brazil\\
}

\date{\today}

\begin{abstract}
Simulations of irreversible growth of extended (fractal and square) islands
with critical island sizes $i=1$ and $2$ are performed in broad ranges of coverage $\theta$ and
diffusion-to-deposition ratios $R$
in order to investigate scaling of island size and capture zone area distributions (ISD, CZD).
Large $\theta$ and small $R$
lead to a crossover from the CZD predicted by the theory of Pimpinelli and Einstein (PE),
with Gaussian right tail, to CZD with simple exponential decays. The corresponding ISD also cross over
from Gaussian or faster decays to simple exponential ones. For fractal islands, these features are
explained by changes in the island growth kinetics,
from a competition for capture of diffusing adatoms (PE scaling) to
aggregation of adatoms with effectively irrelevant diffusion,
which is characteristic of random sequential adsorption (RSA) without surface diffusion.
This interpretation is confirmed by studying the crossover with similar CZ areas (of order
$100$ sites) in a model with freezing of diffusing adatoms that corresponds to $i=0$.
For square islands, deviations from PE predictions appear for coverages near $\theta =0.2$ and
are mainly related to island coalescence. Our results show that the range of applicability of
the PE theory is narrow, thus observing the predicted Gaussian tail of CZD may be difficult
in real systems.
\end{abstract}

\maketitle

\section{Introduction}
\label{intro}

The theoretical study of thin film and multilayer growth attracted much interest
in the last decades motivated by the increasing number of experimental techniques and
applications \cite{etb,pimpinelli,frontiers}. Film morphology is strongly connected to the initial
stages of its growth, where islands of various shapes may be formed. The submonolayer
regime, in which a single incomplete layer is being formed, was modeled by various authors
and is still the focus of much debate \cite{etb,venables,ratsch,af1995,mb,pe,shi,li,pereply,submonotiago}.
The theoretical approaches usually try
to explain the island size distributions (ISD) and the capture zone distributions (CZD),
with a capture zone (CZ) defined as the area in which a diffusing adatom is more 
likely to attach to a given island than to any other one.
The models were originally proposed for atomic epitaxy and recently extended to growth of
other materials, such as 
organic molecule islands, colloidal epitaxy, and graphene epitaxy 
\cite{mulheranPRL2008,clancy2006,ganapathy,lloyd,conrad,groce}.

For irreversible island growth,
the most recent advance in the field is the theory of Pimpinelli and Einstein (PE) \cite{pe},
which proposed that CZD are described by the Wigner surmise (WS) from random matrix theory \cite{rmt}.
The basis of PE theory is the competition of CZs for aggregation of diffusing adatoms.
Deviations in the peaks of simulated CZD and the WS were reported for point islands on surfaces \cite{shi}.
However, a recent work showed excellent collapse of most CZD with the WS after suitable rescaling,
for point, fractal, and square islands, and showed that
all CZD had the universal Gaussian decay predicted by PE \cite{submonotiago}. Indeed, that theory
obtains the WS from a Langevin equation, which is expected to apply only
for large CZs. Moreover, the tail of the ISD was predicted by a scaling
approach that assumes the Gaussian decay of CZD and accounts for the island shape \cite{submonotiago}.

On the other hand, failure of the predicted decay of CZD was observed for point islands in
one-dimensional lattices, where a decay as $exp{\left( -x^3\right)}$ is found
\cite{gonzalez2011a,grinfeld,oneill}. This decay is also predicted analytically in two dimensions
under a circular-cell approximation for CZs \cite{gonzalez2011b}.
For the more realistic cases of extended islands, there are additional reasons to expect limitations
of the PE theory. First, as $\theta$ increases, the size of island-free regions decreases,
enhancing the correlations between stable islands and consequently changing their competition dynamics.
Moreover, if the diffusion-to-deposition ratio $R\equiv D/F$ is small,
the diffusion lengths of the free adatoms are small,
facilitating formation of new islands instead of capture by the existing ones.

This work is devoted to study the crossover from PE submonolayer growth (Gaussian CZD
and related forms of ISD) in those conditions.

Simulations of irreversible growth of fractal and square islands
with critical island nucleus of sizes $i=1$ and $2$ are performed with broad ranges of $\theta$ and $R$.
There is a variety of shapes of CZD and ISD for small CZ areas and small islands (left tails of
CZD and ISD). On the other hand, the right tails of CZD and ISD always show a crossover to simple
exponential decay for small $R$ or large $\theta$. In some cases, it restricts
the PE prediction to a narrow ranges of those variables.

Those deviations are related to the decrease of the size of free regions between the islands,
which become very close and strongly correlated, ruling out the (mean-field) PE theory
that neglects island correlations.
For small $R$, the exponential tail is an effect of small diffusion lengths of adatoms before aggregation,
typical of random distributions of non-mobile adatoms,
which is the random sequential adsorption (RSA) problem \cite{privman,evansrev2}.
As $\theta$ increases, for any $R$, fractal islands become large, thus most of the atoms are deposited
near their branches and rapidly aggregated. Again, adatom diffusion lengths are small and
RSA scaling appears.
A CZ area reduction to less than $100$ sites is characteristic of this crossover region.
This is confirmed by results of a model with $i=0$ where diffusing adatoms may
stop moving without aggregating to an existing island (adatom freezing) \cite{af1995}.
The crossover in square island models as $\theta$ increases
and the islands begin to coalesce is interpreted along the same lines.

The study of ISD and CZD is also important for experimental work \cite{ruiz,wu,shiqin,zheng},
particularly to determine the critical island size that reveals basic features of the aggregation
processes. At this point, we note that recent works compared CZD and ISD with fitting curves
including the WS, such as para-sexiphenyl island growth on different substrates \cite{lorbek,potocar},
$Cu$ deposition with impurities \cite{sathiyanarayanan}, pentacene island growth with
impurities \cite{conrad}, $InAs$ quantum dot
growth on $GaAs$ \cite{arciprete2010}, and $C_{60}$ deposition  on $SiO_2$ films \cite{groce}.

The rest of this work is organized as follows. In Sec. \ref{models}, we present the growth models
and summarize previous theoretical results. In Sec. \ref{fractal}, we discuss the scaling of CZD
and ISD of fractal islands for broad ranges of coverage and diffusion-to-deposition ratio, with special
emphasis
on deviations from PE predictions. In Sec. \ref{square}, the discussion is extended to square islands
models. In Sec. \ref{freezing}, we analyze a model where diffusing adatoms may freeze, which presents
a crossover from regimes with relevant and irrelevant adatom diffusion. Sec. \ref{conclusion}
summarizes our results and presents our conclusions.

\section{Model definition and theoretical approaches}
\label{models}

We consider models of atom deposition and diffusion on a square lattice with irreversible
aggregation to islands larger than a critical size $i$. Atoms incide at randomly chosen
lattice sites
with rate $F$ (number of deposited atoms per unit time per lattice site). An atom adsorbs at the
incidence site
only if it is empty, thus growth is restricted to a single layer.
An adatom diffuses with coefficient $D$ (number of random steps to neighboring sites per unit time)
if it is not aggregated to a stable island. The critical
nucleus $i$ is the maximum size of a non-stable island, i. e. an island from which atoms can detach.
Islands of size $i+1$ or larger are stable.

The diffusion-to-deposition ratio is defined as $R\equiv D/F$.
The coverage $\theta$ is defined as the ratio between the number of adsorbed atoms and
the number of lattice sites.

In fractal island models, each atom permanently aggregates at the site where it collides with a
stable island. Since there is no relaxation of the aggregated atoms, branched shapes are produced,
resembling those of diffusion-limited aggregation \cite{witten}.

In square island models, after aggregation to a stable island, the atom instantaneously relaxes to a position
that preserves the compact (square) shape \cite{sqevans}. However, when two neighboring islands
coallesce, they evolve independently: an atom aggregated at the border of one of the
coallesced islands relaxes to a site in the border of that island to preserve its square shape.

In the case of point island models, all atoms of an island aggregate on a single lattice site.
However, this case is not studied in this work, since our aim is to study features of extended
islands.

It is reasonable to assume that the probability density of CZ area $x$ follows the
scaling form
\begin{equation}
P(x)=\frac{1}{\langle x\rangle}
f{\left( \frac{x}{\langle x\rangle}\right)} ,
\label{scalingtrad}
\end{equation}
where $f$ is a scaling function.
An equivalent scaling form applies to the density of islands of size $s$, $Q(s)$.
A recent work \cite{submonotiago} proposed the scaling of CZD with the variance
$\sigma_x\equiv {\overline{{\left( x-\langle x\rangle\right)}^2}}^{1/2}$ as
\begin{equation}
P(x) = \frac{1}{\sigma_x} g{\left( \frac{x-\langle x\rangle}{\sigma_x}\right)} .
\label{scalingsigma}
\end{equation}
This procedure is inspired in the successful application to roughness distributions of
thin film growth models \cite{intrinsic,graos}.
An equivalent form also applies to $Q(s)$.

In 1995, Amar and Family (AF) proposed the empirical formula for the ISD \cite{af1995}
\begin{equation}
f_i\left( u\right) = C_i u^i \exp{\left( -ia_i u^{1/a_i}\right)},
\label{af}
\end{equation}
where $u\equiv s/\langle s\rangle$ and $C_i$ and $a_i$ are normalization constants.
This formula is widely used to fit experimental data \cite{ruiz,wu,shiqin,zheng}.
Subsequently, Mulheran and Blackman (MB) proposed to relate the ISD to
distributions of areas of Voronoi polygons and provided results close to
those of point island models \cite{mb}.
In all those works, comparison with simulation or experimental data focused on the peaks
of the distributions.

A recent work by K\"orner et al \cite{korner}
proposed mean-field rate equations accounting for coverage-dependent capture numbers
and obtained good agreement with the peaks and the tails of the ISD
for fractal islands. However, the theoretical ISD were obtained by numerical integration of
those equations, thus the tail decays were not analyzed.

Recently, a significant advance was the proposal \cite{pe}
that CZD are described by the WS
\begin{equation}
P_\beta(z) = a_\beta z^{\beta} \exp{\left( -b_\beta z^2\right)} ,
\label{ws}
\end{equation}
where $z\equiv x/\left\langle x\right\rangle $, $\beta = \frac{2}{d}\left( i+1\right)$,
$d$ is the substrate dimension ($d=2$ in
the present work) and the parameters $a_\beta$ and $b_\beta$ are determined by normalization conditions.
The PE theory is justified by a phenomenological argument that the CZD
can be extracted from a Langevin equation representing the competition of neighboring islands
for adatom aggregation.

Ref. \protect\cite{submonotiago} showed that the CZD scaling using Eq. (\ref{scalingsigma})
agrees with the WS of Eq. (\ref{ws}) for large values of $R$ (typically $R\geq {10}^7$)
and small coverages, in point, fractal, and square island models.
In all cases, the Gaussian right tail predicted by PE [Eq. (\ref{ws})] was present.
For point and square islands, some deviations in the left tails and in the peaks of the CZD appeared.
However, deviations for small islands and small CZs are expected because the continuous approach
of PE applies to large island and large CZ kinetics.

Ref. \protect\cite{submonotiago} considered $\beta=i+1$ in the WS, as originally suggested \cite{pe}.
The successful comparison is probably due to the adopted rescaling and
contrasts to the proposal $\beta=i+2$ of Refs. \protect\cite{shi,li,pereply}.
The Gaussian decay also differs from the $exp{\left( -x^3\right)}$ decay
predicted in Ref. \protect\cite{gonzalez2011b} for point islands, probably due to the use of a
circular-cell approximation for CZs in that work.

A scaling approach predicts the decay of the right tails of ISD
from the Gaussian tail of the CZD \cite{submonotiago}. For fractal islands, the decay is
$\exp{\left( -s^{4/D_F}\right)} \approx \exp{\left( -s^{2.36}\right)}$, where 
$D_F\approx 1.694$ \cite{meakin83} is the fractal dimension of DLA clusters;
for square islands, the ISD decay is also Gaussian. These results also agree with simulation
data for large $R$ and small coverages \cite{submonotiago}.

PE theory applies to systems where distant and large islands compete for aggregation of
diffusing adatoms. The opposite situation is that of totally non-mobile adatoms, which is
the random sequential adsorption (RSA) problem without diffusion \cite{privman,evansrev2}.
In this case, atoms irreversibly stick to the site where they are adsorbed.
ISD and CZD have simple exponential decays \cite{stauffer} as
\begin{equation}
Q(s) \sim \exp{(-s/{\langle s\rangle})} .
\label{expdecay}
\end{equation}

\section{Distributions for fractal islands}
\label{fractal}

\subsection{Simulation results}
\label{fractalsimulation}

We performed simulations in lattices of very large lateral size, typically $L=2048$,
with $R$ ranging between ${10}^4$ and ${10}^{9}$, coverages up to $\theta = 0.4$, and critical
nucleus $i=1$ and $i=2$. For $i=2$, no energy barrier is considered for the dissociation of
islands with two atoms.

\begin{figure}[!t]
\includegraphics[width=8cm]{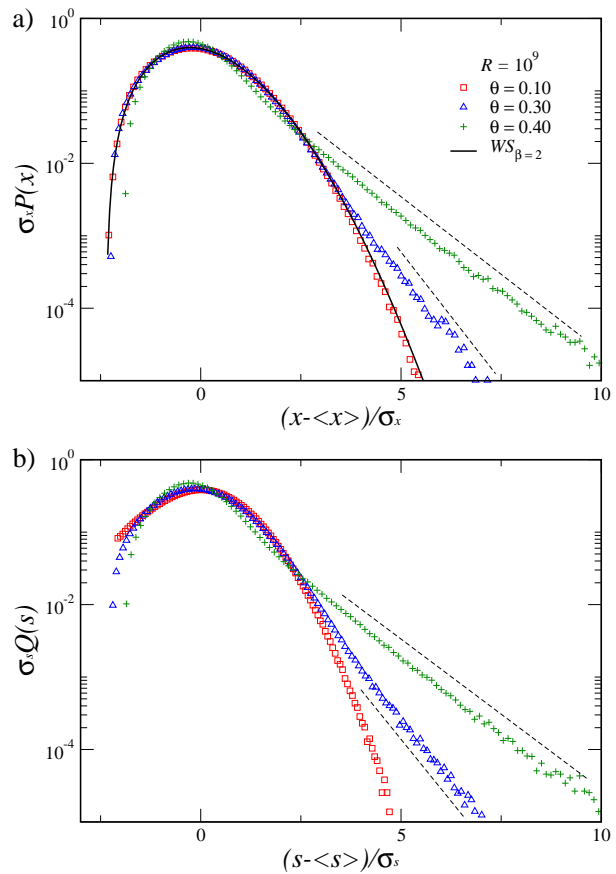}
\caption{(Color online) Scaled CZD (a) and ISD (b) for fractal islands with $i=1$, $R=10^{9}$ and several coverages. The solid line is the WS with $\beta=2$. Dashed lines in all plots are guides to the eye.}
\label{fig1}
\end{figure}

In Fig. 1a, we show scaled CZD for $R={10}^9$ and various coverages, to be compared with the WS.
The data for small coverage ($\theta = 0.1$ in Fig. 1a) is well fit by the WS, as previously
shown in Ref. \protect\cite{submonotiago}. However, deviations
clearly appear for the largest coverages ($\theta =0.3$ and $0.4$) and
a simple exponential decay of the right tail is found, in contrast to the Gaussian decay of the WS.

In Fig. 1b, we show ISD for the same set of parameters. For small coverages, the right tail decay is
slightly faster than a Gaussian, as discussed in Sec. \ref{models} (see also Ref. \protect\cite{submonotiago}).
However, it also crosses over to simple exponential decay for the largest coverages ($\theta =0.3$ and $0.4$).

The same crossover is found as $R$ decreases, for fixed $\theta$. In Fig. 2a, we show
CZD for $\theta=0.10$ and several values of $R$. The CZD for $R={10}^8$ agrees very well with the WS.
However, the CZD for smaller values of $R$ deviate from the WS and these deviations are enhanced
for decreasing $R$. The simple exponential decay is found for large CZs for $R \lesssim {10}^5$.
Despite the deviation in the tails, for $R={10}^6$, the peak of the CZD still shows reasonable agreement
with the WS (inset of Fig. 2a). For smaller $R$, the peaks also deviate from the WS,
compensating the large difference in the tails.

In Fig. 2b, we show CZD for $\theta=0.3$ and several values of $R$. For $R={10}^9$, the CZD
agrees with the WS in two orders of magnitude around the peak [$\sigma_zP\left( z\right) \sim {10}^{-2.5}$
to ${10}^{-0.5}$]. For smaller values of $R$, the fit covers a smaller region of the scaled CZD.
The simple exponential decay of the right tail is observed for large CZs in all cases.
The corresponding ISD are shown in Fig. 2c and also have that decay.

\begin{figure}[!t]
\includegraphics[width=8cm]{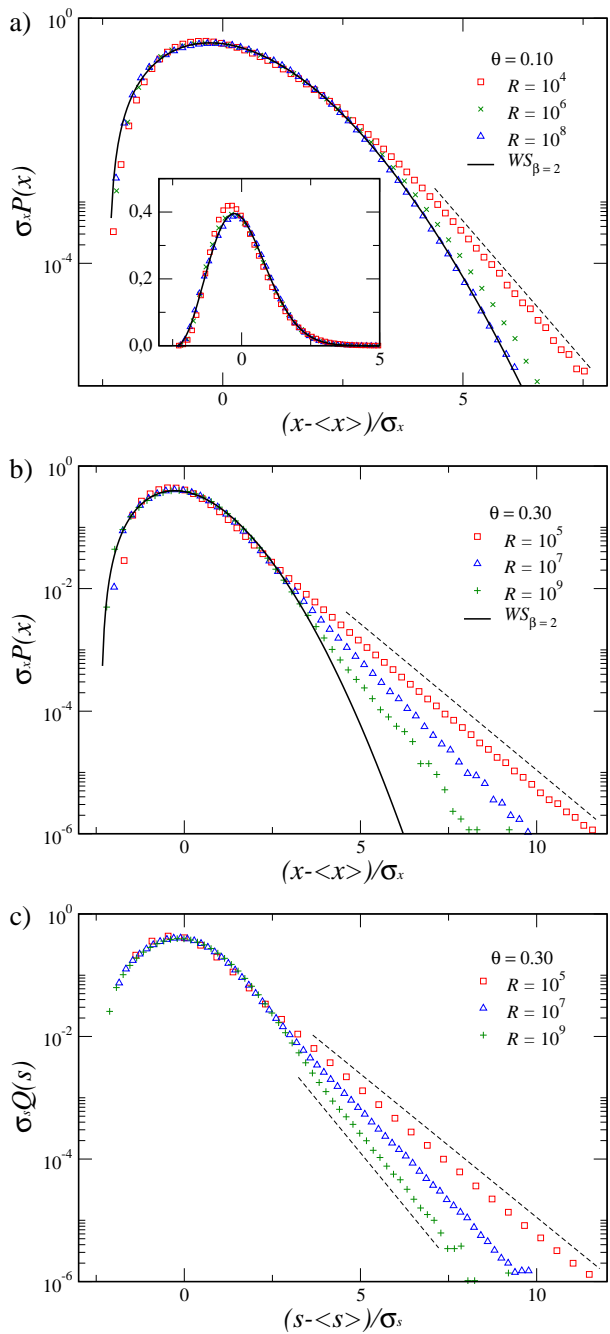}
\caption{(Color online) (a) Scaled CZD for fractal islands with $i=1$, $\theta=0.10$ 
and several ratios $R$. The inset in (a) shows the same data of the main plot in linear-linear scale.
Scaled CZD (b) and ISD (c) for fractal islands with $i=1$, $\theta=0.30$
and several ratios $R$. The solid line is the WS with $\beta=2$. Dashed lines in all plots are guides to the eye.}
\label{fig2}
\end{figure}

Note that coverage $\theta=0.3$ is typical of many experimental works because islands become
sufficiently large and facilitate microscopy investigation with a focus on island statistics.
Moreover, $R \lesssim {10}^5$ is typical of low temperatures, where island relaxation is
difficult and the fractal island model is realistic.
These features suggest that experimental observation of PE predictions [particularly the
Gaussian tail of CZD and the related form of the ISD (Sec. \ref{models})] will be difficult
when large fractal islands are grown.

Fig. 3 shows images of the submonolayers grown with $R={10}^7$ and $R={10}^9$.
For $\theta=0.1$ and $R={10}^9$, the clusters are distant from their neighbors,
i. e. there are large areas between them. The PE theory actually applies
to this situation \cite{submonotiago}. However, for $R={10}^7$ and the same coverage, islands are much
smaller, the distance between them decreases and the free areas around them are smaller.
Under these conditions, the hypothesis of PE theory begin to fail.
For $\theta=0.1$ and $R={10}^6$, these effects are enhanced, which explains deviations of the CZD from the WS,
as shown in  Fig. 2a.

\subsection{Interpretation of results}
\label{fractalinterpretation}

First we consider the deviations from PE scaling observed for small values of $R$.

In this case, the average adatom diffusion length $\langle l_D\rangle$ is very small, as
well as its average diffusing time $\langle t_D\rangle$. Island growth mainly
proceeds by incidence of new atoms very close to existing islands or very close to a diffusing
adatom. If the lattice size is rescaled by $\langle l_D\rangle$ and time is rescaled by
$\langle t_D\rangle$, then
the islands of the rescaled lattice will grow by attachment of atoms deposited at neighboring sites,
i. e. non-diffusion atoms.
This leads to distributions with right tails (large islands) similar to RSA (Eq. \ref{expdecay}).

Another important aspect to explain the deviations from PE scaling for small $R$ is the small average
island size and small average CZ area, since that theory follows from continuous approaches suitable
for large CZs.

This interpretation helps to quantify the deviation of the DCZ from the WS as $R$ decreases.

\begin{figure*}[t]
\centering
\includegraphics[width=15cm]{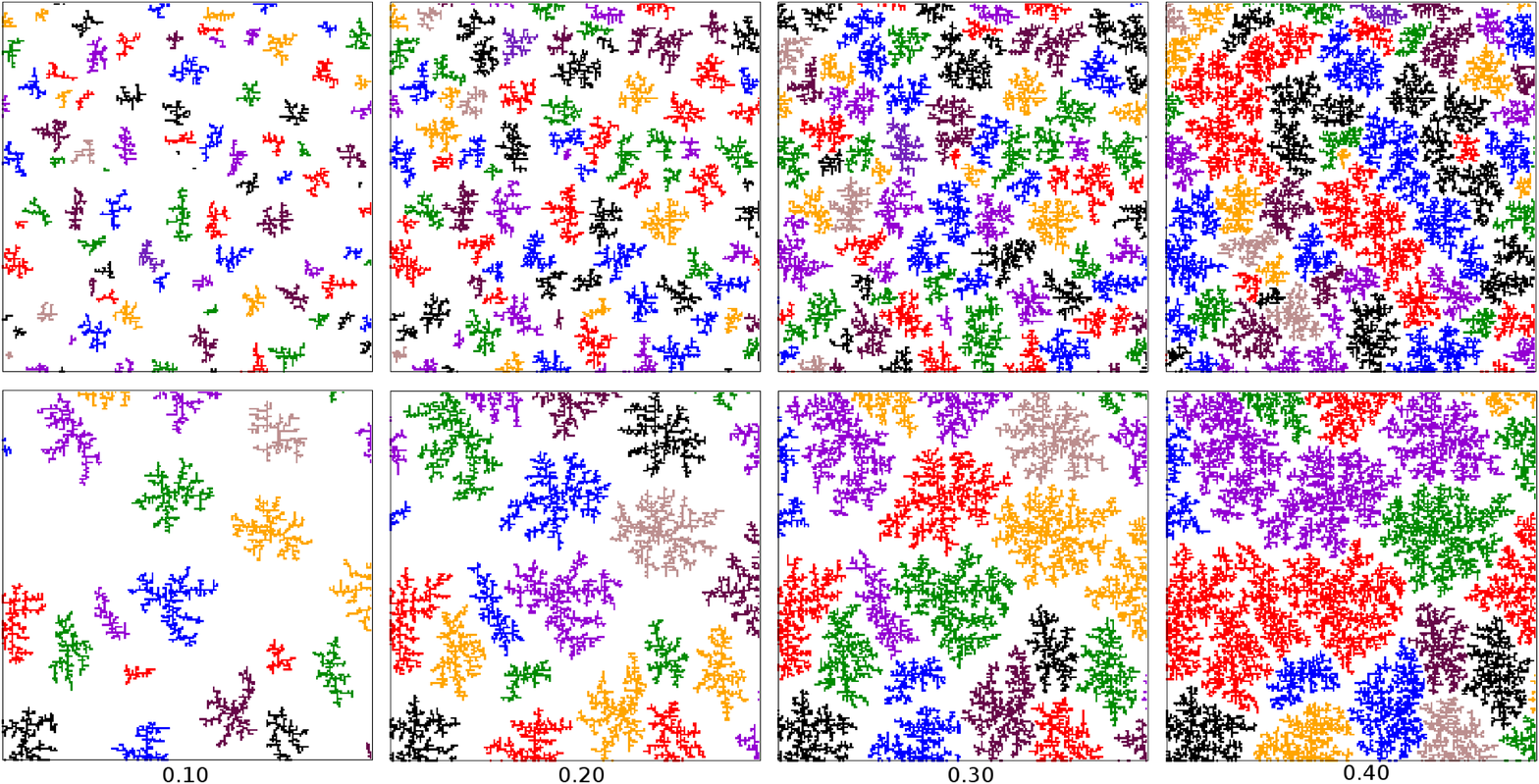}
\caption{(Color online) Fractal islands for several coverages with $R=10^7$ (top) and $R=10^9$ (bottom).
The panels of lateral size $200$ sites are cut from a system of size $L=2048$.}
\label{fig3}
\end{figure*}

\begin{figure}[!h]
\centering
\includegraphics[width=8cm]{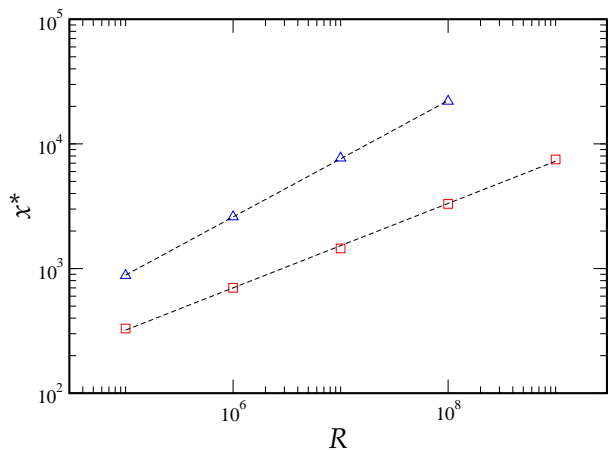}
\caption{(Color online) CZ crossover area $x^*$ as a function of $R$, for $\theta = 0.30$ and critical
islands sizes $i=1$ (squares, red) and $i=2$ (triangle, blue).
The error bars are of the same order of the point size. Dashed lines are least squares fits of the data.}
\label{fig4}
\end{figure}

The deviation occurs for a CZ area $x^*$ in which the WS decay
$\exp{\left[ -{\left( x/{\langle x\rangle}\right)}^2\right] }$ (Eqs. \ref{scalingtrad} and
\ref{ws}), matches the simple exponential decay of Eq. \ref{expdecay}.
Thus $x^* \sim {\langle x\rangle}$. In the crossover, islands are large and very close
(see e. g. Fig. 3), branched but not fractal, thus the CZ area $x^*$ is of the same order of
the average island size.
The dependence of this size on $R$ and $\theta$ in the steady state of island
growth is predicted by rate equation theory \cite{etb,venables1,venables2,stoyanov}
as $\langle s \rangle \sim R^{i/\left( i+2\right)} \theta^{-\left( i+1\right) /\left( i+2\right)}$.
This gives
\begin{equation}
x^* \sim R^{i/\left( i+2\right)} \theta^{-\left( i+1\right) /\left( i+2\right)} .
\label{scalingxast}
\end{equation}

The crossover area was determined in the simulated CZD as the point in which the tail changes
its concativity in a log-linear plot. Error bars were estimated as the whole range of $x$
that covers the crossover region. 
Fig. 4 shows the scaling of $x^*$ with $R$, for fixed $\theta=0.3$, with linear fits giving scaling
exponents $\chi=0.33 \pm 0.02$ for $i=1$ and $\chi=0.47 \pm 0.03$ for $i=2$.
This result is in good agreement with the predictions $x^*\sim R^{1/3}$ and $x^*\sim R^{1/2}$ of
Eq. (\ref{scalingxast}) for $i=1$ and $2$, respectively.
The scaling of the average island size at the crossover, $s^*$, is analogous to $x^*$,
and was also confirmed numerically (results not shown).

Another interesting conclusion from this quantitative analysis is that the rate equation theory
is still a reliable approximation for average quantities, despite the changes in the main kinetic
mechanisms determining the evolution of large CZs and large islands (responsible for the crossover
in the right tails of ISD and CZD).

Now we consider the deviations from PE behavior with increasing coverage.
For instance, Fig. 3 shows that the islands are very close (so, strongly correlated) for $R={10}^7$ and $\theta=0.2$.
They grow mainly by random aggregation of mass deposited near its branches, and not by capturing
diffusing atoms that incided in island-free areas. Thus, there is significant increase in the
widths of island branches when $\theta$ evolves to $0.3$ and $0.4$, instead of increase of
the length of the branches (in other words, islands become less ramified).


This situation may also be viewed as a case of small average adatom diffusion
lengths because the free area available is small. Again, rescaling of the
lattice by $\langle l_D\rangle$ and of the time by $\langle t_D\rangle$ indicates that most atoms
will aggregate immediately after incidence. This corresponds to RSA scaling.

It is worth to recall that an attempt to deposit an atom on an already occupied site is rejected
in our models, and a new site is randomly chosen. However, in many simulation works, that atom is
randomly aggregated in the island periphery. For low coverages, this condition is irrelevant, but
for large coverages the second rule increases the number of static adatoms aggregated to islands.
Thus, the crossover effects would be enhanced if this (more realistic) mechanism is used.

\section{Distributions for square islands}
\label{square}

We performed simulations of the square island model in the same range of parameters of the fractal
island simulations.

In Figs. 5a and 5b, we show CZD for $R={10}^9$ and various coverages, in logarithmic and linear plots,
respectively.
The data for small coverage ($\theta =0.1$) is well fitted by the WS, but deviations
are large for $\theta =0.25$ and larger coverages. Again, the Gaussian decay of the right tail
of the CZD (PE scaling) crosses over to simple exponential decay as $\theta$ increases.

\begin{figure}[!t]
\includegraphics[width=8cm]{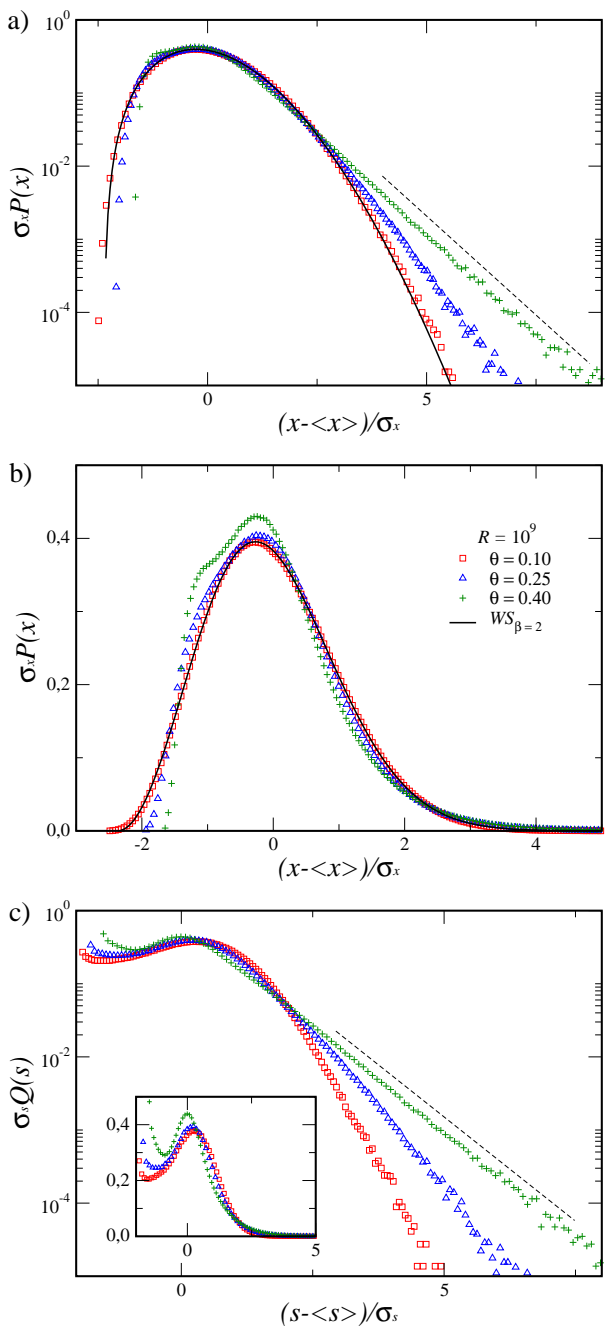}
\caption{(Color online) Scaled CZD (a,b) and ISD (c) for square islands  with $i=1$, $R=10^{9}$
and several coverages. The solid line in (a) and (b) is the WS with $\beta=2$. Dashed line in (a), (c) are guides to the eye.}
\label{fig5}
\end{figure}

In Fig. 5c, we show ISD for the same set of parameters, showing the crossover from the
Gaussian decay (small coverage) to the simple exponential (large coverage). The inset of Fig. 5c
highlights the small island densities.

Important differences from the fractal island case can be noted in the distributions of Figs. 5a-c.

The first one
is the presence of many small islands for all coverages, which was already noted in Ref.
\protect\cite{barteltSS1993}. For $\theta=0.4$, the density of small islands
($s=2, 3, 4$) is comparable to the density of islands of the average size ($s\approx 1100$).
This certainly affects all results from continuous approaches, which is the case of
the PE theory, and justifyes the deviations between the CZD and the WS for relatively small coverages.

Another difference is the formation of a shoulder in the CZD as $\theta$ increases (Fig. 5b).
This is also related to the presence of small islands. Despite their small size, their CZ areas are
not small; instead, they are only slightly smaller than the most probable CZ areas (peaks of CZD).

Results for fixed coverage and decreasing $R$ show the same crossover of the fractal islands
(Figs. 2a and 2b). However, the simple exponential decay of the right tails (of CZD and ISD)
appears for larger values of $R$, i. e. the Gaussian CZD is observed in a narrower range for
square islands.

Fig. 6 shows images of the submonolayers grown with $R={10}^7$ and $R={10}^9$. For any coverage,
the free areas around the square islands are larger than those around the fractal islands.
Thus, one should expect PE theory to apply for larger coverages with square islands.
However, the opposite occurs: deviations of the CZD from the WS appear for smaller coverages
when compared to fractal islands.

The formation of small islands explains the deviations in the left tail and in the peak of CZD.
On the other hand, the change in the right tail scaling (from Gaussian to simple exponential) is not
explained by this feature, but is related to island coalescence.
For $\theta=0.2$, only two-island coalescence can be seen in Fig. 6 and is significant
for $R={10}^7$. For $\theta =0.4$, coalescence of many islands is noticeable for both
values of $R$. Now suppose that the system
is rescaled by a factor $\langle s\rangle$, so that an average island of the original system
is converted to a single occupied site. After rescaling, the largest islands are formed
by coalescence of some randomly distributed sites, which were the coalesced islands in the original
system. Thus, ISD and CZD have right tails as in random distribution of occupied sites, i. e. RSA.

\section{Model with diffusion and freezing ($i=0$ case)}
\label{freezing}

This model was introduced by Amar and Family \cite{af1995} to represent growth with critical
island size $i=0$ incorporating diffusion effects. That model is different from the
random sequential adsorption (RSA), in which atoms
irreversibly stick to the site where they are adsorbed, without diffusion \cite{privman,evansrev2}.

\begin{figure*}[!t]
\centering
\includegraphics[width=15cm]{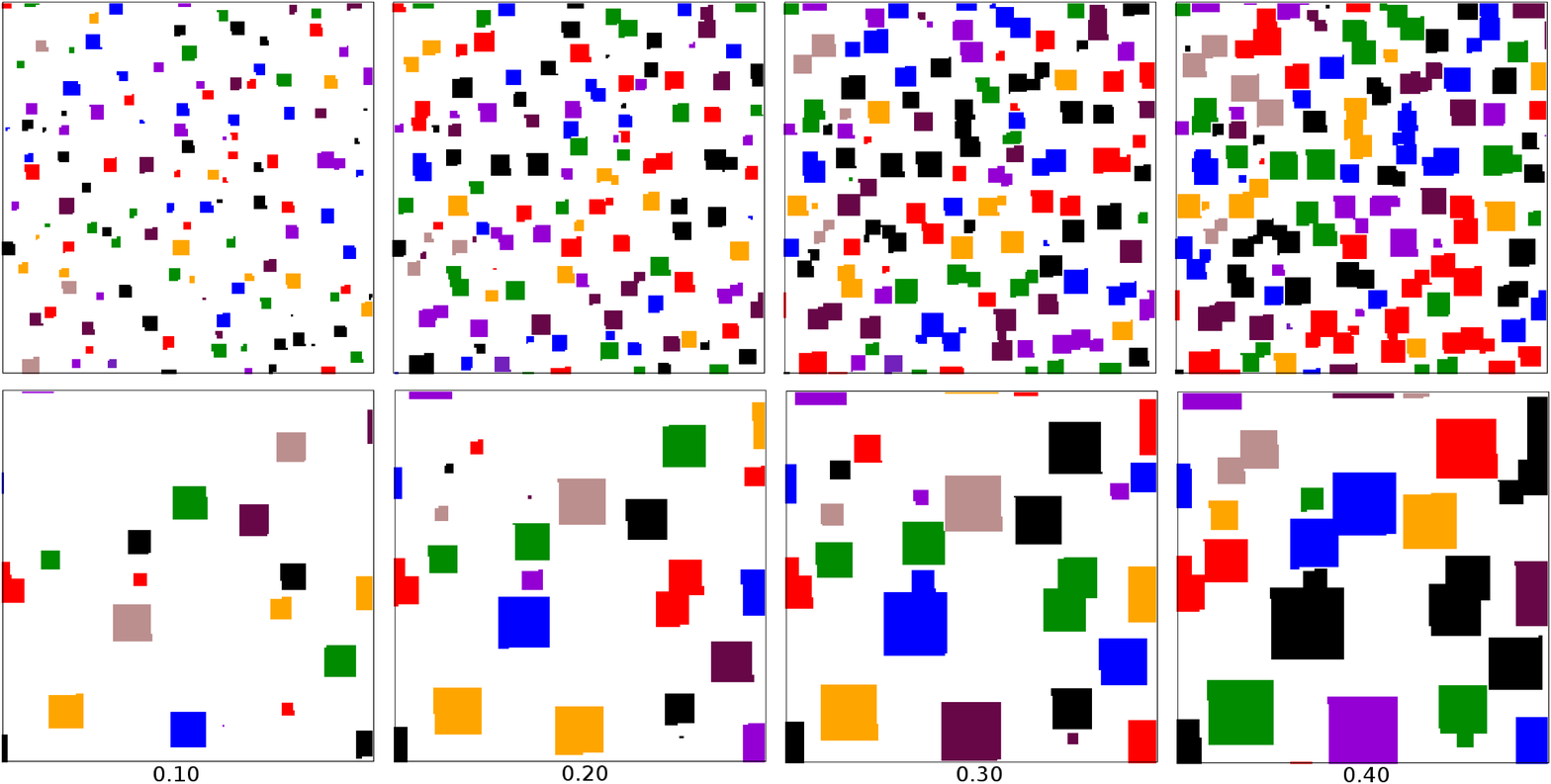}
\caption{(Color online) Square islands for several coverages with $R=10^7$ (top) and $R=10^9$ (bottom).
The panels of lateral size $200$ sites are cut from a system of size $L=2048$.}
\label{fig6}
\end{figure*}

In the model, the adatoms diffuse with coefficient $D$
and stop moving (freeze and nucleate a new stable island) with rate $R_s = rD$, with $r\leq 1$.
The freezing rate depends on an extra activation
energy $E_F$ as $r=\exp{\left(- E_F/k_BT\right)}$, which may represent the effect of surfactants
or impurities \cite{rosenfeld,chambliss}. Adatoms also stop moving when they become
nearest neighbors of stable islands, producing islands with fractal shape.
However, an adatom does not stop moving when it meets one
or more diffusion adatoms, which is the main difference from the models with $i>0$.

We performed simulations of this model in lattices of lateral size $L=2048$,
with $R$ between ${10}^9$ and ${10}^{11}$, coverages up to $\theta = 0.3$, and $r$ ranging
from ${10}^{-2}$ to ${10}^{-5}$.

In Figs. 7a and 7b we show scaled CZD and ISD, respectively, for different
freezing rates and fixed $R$ and $\theta$. For $r=10^{-5}$, the CZD shows a reasonable
agreement with the PE theory for $i=0$ (WS with $\beta=1$). However, a large deviation is
observed for $r=10^{-3}$, where the tail decay is slower than a Gaussian (see inset of Fig. 7a).
The ISD have simple exponential decay for $r=10^{-3}$ and large $s$, again typical of RSA.

We notice that, in this model, islands are produced only by adatoms that stop moving. Thus, the
number of islands increases with the rate $r$, decreasing the CZ areas and the size of the free
regions between the islands.

The crossover in CZD can be explained by a comparison of CZ areas with the
model with $i=1$ (Sec. \ref{fractal}).

The average CZ area for $r=10^{-5}$ and $\theta =0.1$ is $\langle x\rangle \approx 880$.
Comparing with fractal islands grown with $i=1$ and the same coverage, we note that this value
is between the CZ area for $R=10^7$ ($\langle x\rangle \approx 500$) and for $R=10^8$
($\langle x\rangle \approx 1200$). Both CZD agree with the WS with $\beta = 2$, as shown in Ref.
\protect\cite{submonotiago} (see also Fig. 2a for $R=10^8$).
The case $R=10^7$ is also illustrated in Fig. 3 and show
large areas between the islands.

\begin{figure}[!t]
\centering
\includegraphics[width=8cm]{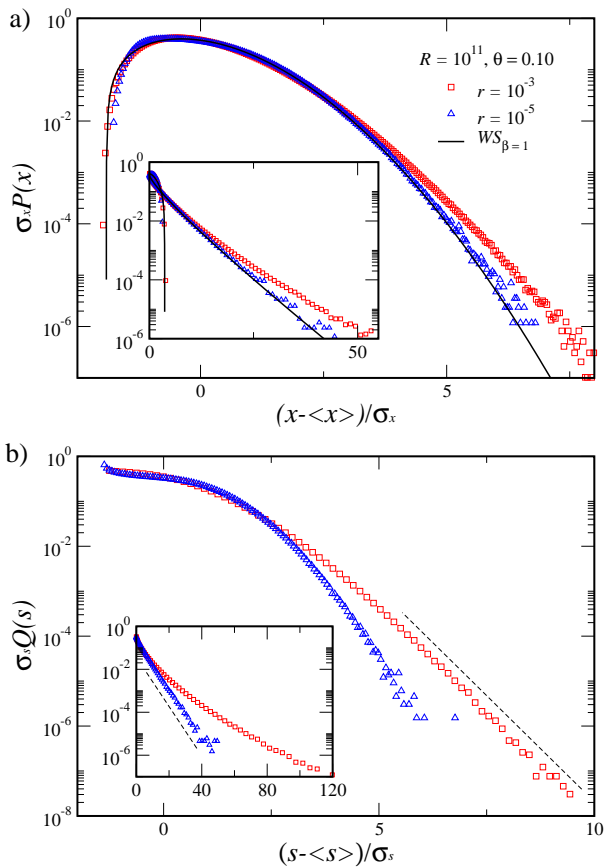}
\caption{(Color online) Scaled CZD (a) and ISD (b) for $i=0$ model with $R=10^{11}$, $\theta=0.10$ and
$r=10^{-3}$ (squares, red) and $r=10^{-5}$ (triangles, blue). The solid line is the WS with $\beta=1$.
The insets show the same distributions with abscissas
$\left[ \frac{(z-\left\langle z\right\rangle )}{\sigma_{z}} \right]^2$ in (a) and
$\left[ \frac{(s-\left\langle s\right\rangle )}{\sigma_{s}} \right]^{2.2}$ in (b).}
\label{fig7}
\end{figure}

On the other hand, the average CZ area for $r=10^{-3}$ and $\theta =0.1$ is much smaller:
$\langle x\rangle \approx 88$. This value is close to the average CZ area for fractal islands
with $i=1$, $\theta = 0.1$, and $R=10^4$: $\langle x\rangle \approx 64$
(for $R=10^5$, we find $\langle x\rangle \approx 120$).
The corresponding CZD is shown in Fig. 2a and has significant deviation from the WS with $\beta = 2$.

The deviations from PE scaling for small $R$ in fractal islands
are related to the decrease of the average CZ area and the corresponding decrease
in the available area between the islands. The crossover occurs for an average CZ area near $100$
lattice sites, which apparently independs on the value of $i$, reinforcing the geometric
argument. 

The case $r=1$ is equivalent to the RSA, since adatoms freeze with
the same rate that they move to neighboring sites. Consequently, our results consistently
illustrate a crossover from a regime of island growth from diffusing adatoms
to a regime of island formation by approximately static adatoms.

\section{Conclusion}
\label{conclusion}

We simulated irreversible growth of fractal and square islands with critical island size $i=1$ and $2$,
for several values of diffusion-to-deposition ratio $R$ and coverage $\theta$, and simulated a
model with $i=0$, with adatom diffusion and freezing. ISD and CZD were calculated and the
crossover from PE theory predictions to different distribution shapes was analyzed, with a focus
on the right tail decays.

For fractal islands, small $\theta$ and large values of $R$ lead to Gaussian CZD, which is
consistent with the PE theory. ISD have left tails slightly faster than Gaussian.
However, as the island density increases (due to increasing $\theta$ or decreasing $R$),
the area between neighboring islands becomes very small.
This strong correlation between islands leads to a crossover to simple exponential decays of CZD and ISD.
For small $R$ or large $\theta$, the deposited adatoms have very small diffusion lengths before
aggregating to a stable island, which is interpreted as a crossover to the problem of
random distributions of static atoms (RSA problem). For fixed coverage, estimates of the island sizes
and CZ areas at the crossover have $R$-scaling consistent with rate equation theory for $i=1$ and $i=2$.

For square islands, small $\theta$ and large values of $R$ lead to Gaussian CZD and ISD.
Compared to the fractal island case, the crossover is observed for smaller values of $\theta$
and larger values of $R$, despite the larger area between neighboring islands. In the square island case,
the deviations from PE predictions is partially atributed to the presence of many small islands,
whose densities are of the same order of the (much larger) average islands.
However, the crossover from Gaussian to simple exponential decay in CZD and ISD is mainly related to 
coalescence of islands, since large islands and large CZs are effectively formed by
random distribution of (static) rescaled islands.

In the model with diffusion and freezing, island shape is also fractal.
For small values of the freezing rate $r$, islands are separated by large areas and the CZD obeys
the PE theory for $i=0$. As $r$ increases, the area between the islands become very small, so that
CZD and ISD get a simple exponential decay, similar to the fractal islands with $i=1$.

The Gaussian tail of CZD and related forms of ISD are found only for small
coverages (usually $\theta < 0.3$) and for large values of $R$, thus their experimental
observation is probably difficult.
Moreover, large values of $R$ are characteristic of high temperatures, where the hypothesis of
a critical (and small) island size may fail. This strongly suggests extensions of the theoretical
and simulational study to reversible island growth and to systems with cluster diffusion,
which were focus of recent work \cite{kryukov}.

\acknowledgments

The authors acknowledge support from CNPq, FAPEMIG and FAPERJ (Brazilian agencies).


\newpage


\begin{references}

\bibitem{etb} J.W. Evans, P. A Thiel, and M. C. Bartelt, Surf. Sci. 
Rep. {\bf 61}, 1 (2006).

\bibitem{pimpinelli}
A. Pimpinelli and J. Villain, {\it Physics of Crystal Growth} (Cambridge
University Press, Cambridge, England, 1998).

\bibitem{frontiers}
{\it Frontiers in Surface and Interface Science}, edited by Charles
B. Duke and E. Ward Plummer (ÃÂElsevier, Amsterdam, 2002).

\bibitem{venables}
J. A. Venables, {\it Introduction to Surface and Thin Film Processes}
(Cambridge University Press, Cambridge, England, 2000).

\bibitem{ratsch}
C. Ratsch and J. A. Venables, J. Vac. Sci. Technol. A {\bf 21}, S96 (2003).

\bibitem{af1995}
J. G. Amar and F. Family, Phys. Rev. Lett. {\bf 74}, 2066 (1995).

\bibitem{mb}
P. A. Mulheran and J. A. Blackman,  Philos. Mag. Lett. {\bf 72}, 55 (1995);
Phys. Rev. B {\bf 53}, 10261 (1996).

\bibitem{pe}
A. Pimpinelli and T. L. Einstein, Phys. Rev. Lett. {\bf 99}, 226102 (2007).

\bibitem{shi}
F. Shi, Y. Shim, and J. G. Amar, Phys. Rev. E {\bf 79}, 011602 (2009).

\bibitem{li}
M. Li, Y. Han, and J. W. Evans, Phys. Rev. Lett. {\bf 104}, 149601 (2010).

\bibitem{pereply}
A. Pimpinelli and T. L. Einstein, Phys. Rev. Lett. {\bf 104}, 149602 (2010).

\bibitem{submonotiago}
T. J. Oliveira and F. D. A. Aar\~ao Reis, Phys. Rev. B {\bf 83}, 201405(R) (2011).

\bibitem{mulheranPRL2008}
P. A. Mulheran, D. Pellenc, R. A. Bennett, R. J. Green, and M. Sperrin,
Phys. Rev. Lett. {\bf 100}, 068102 (2008).

\bibitem{clancy2006}
D. Choudhary, P. Clancy, R. Shetty, and F. Escobedo, 
Adv. Functional Mater. {\bf 16}, 1768 (2006).

\bibitem{ganapathy}
R. Ganapathy, M. R. Buckley, S. J. Gerbode, and I. Cohen, Science {\bf 327},
445 (2010).

\bibitem{lloyd}
J. H. Lloyd-Williams, B. Monserrat, D. D. Vvedensky, and A. Zangwill,
Phys. Rev. B {\bf 85}, 161402(R) (2012).

\bibitem{conrad}
B. R. Conrad, E. Gomar-Nadal, W. G. Cullen, A. Pimpinelli, T. L. Einstein, and E. D. Williams,
Phys. Rev. B \textbf{77}, 205328 (2008).

\bibitem{groce}
M. A. Groce, B. R. Conrad 1, W. G. Cullen, A. Pimpinelli, E. D. Williams, and T. L. Einstein,
Surf. Sci. {\bf 606}, 53 (2012).

\bibitem{rmt}
M. L. Mehta, {\it Random Matrices} (Academic, New York, 2004), 3rd ed;
T. Guhr et al, Phys. Rep. {\bf 299}, 189 (1998).

\bibitem{gonzalez2011a}
D. L. Gonz\'alez, A. Pimpinelli, and T. L. Einstein, Phys. Rev. E {\bf 84}, 011601 (2011).

\bibitem{grinfeld}
M. Grinfeld, W. Lamb, K. P. O'Neill, and P. A. Mulheran, J. Phys. A: Math. Theor. {\bf 45},
015002 (2011).

\bibitem{oneill}
K. P. O'Neill, M. Grinfeld, W. Lamb, and P. A. Mulheran, Phys. Rev. E {\bf 85},
021601 (2012).

\bibitem{gonzalez2011b}
D. L. Gonz\'alez and T. L. Einstein, Phys. Rev. E {\bf 84}, 051135 (2011).

\bibitem{privman}
V. Privman, Colloids Surf. A {\bf 165}, 231 (2000).

\bibitem{evansrev2} 
J. W. Evans, Rev. Mod. Phys. \textbf{65}, 1281 (1993).

\bibitem{ruiz}
R. Ruiz, B. Nickel, N. Koch, L. C. Feldman, R. F. Haglund, Jr., A. Kahn, F. Family, and G. Scoles,
Phys. Rev. Lett. \textbf{91}, 136102 (2003).

\bibitem{wu}
Y. Wu, T. Toccoli, N. Koch, E. Iacob, A. Pallaoro, P. Rudolf, and S. Iannotta,
Phys. Rev. Lett. \textbf{98}, 076601 (2007).

\bibitem{shiqin}
J. Shi and X. R. Qin, Phys. Rev. B \textbf{78}, 115412 (2008).

\bibitem{zheng}
H. Zheng,  M. H. Xie, H. S. Wu, and Q. K. Xue, Phys. Rev. B \textbf{77}, 045303 (2008).

\bibitem{lorbek}
S. Lorbek, G. Hlawacek, and C. Teichert, Eur. Phys. J. Appl. Phys. {\bf 55}, 23902 (2011).

\bibitem{potocar}
T. Potocar, S. Lorbek, D. Nabok, Q. Shen, L. Tumbek, G. Hlawacek, P. Puschnig, C. Ambrosch-Draxl,
C. Teichert, and A. Winkler, Phys. Rev. B \textbf{83}, 075423 (2011).

\bibitem{sathiyanarayanan}
R. Sathiyanarayanan, A. BH. Hamouda, A. Pimpinelli, and T. L. Einstein,
Phys. Rev. B \textbf{83}, 035424 (2011).

\bibitem{arciprete2010}
F. Arciprete, M. Fanfoni, F. Patella, A. Della Pia, A. Balzarotti, and E. Placidi,
Phys. Rev. B \textbf{81}, 165306 (2010).

\bibitem{witten}
T. A. Witten and L. M. Sander, Phys. Rev. Lett. {\bf 47}, 1400 (1981).

\bibitem{sqevans}
M. C. Bartelt, J. W. Evans, Surf. Sci. 298, 421 (1993).

\bibitem{intrinsic}
T. J. Oliveira and F. D. A. Aar\~ao Reis, Phys. Rev. E {\bf 76}, 061601 (2007).

\bibitem{graos}
T. J. Oliveira and F. D. A. Aar\~ao Reis, J. Appl. Phys. {\bf 101}, 063507 (2007).

\bibitem{korner}
M. K\"orner, M. Einax, and P. Maass, Phys. Rev. B {\bf 82}, 201401(R) (2010).

\bibitem{meakin83}
P. Meakin, Phys. Rev. A {\bf 27}, 604 (1983).

\bibitem{stauffer}
D. Stauffer and A. Aharony, {\it Introduction to Percolation Theory},
2nd. edition (Taylor \& Francis, London/Philadelphia, 1992).

\bibitem{venables1}
J. A. Venables, Phyl. Mag. {\bf 27}, 697 (1973).

\bibitem{venables2}
J. A. Venables, G. D. Spiller, and M. Hanbuchen, Rep. Prog. Phys. {\bf 47}, 399
(1984).

\bibitem{stoyanov}
S. Stoyanov and K. Kashchiev, Curr. Top. Mat. Sci. {\bf 7}, 69 (1981).

\bibitem{barteltSS1993}
M. C. Bartelt and J. W. Evans, Surf. Sci. {\bf 298}, 421 (1993).

\bibitem{rosenfeld}
G. Rosenfeld, R. Servaty, C. Teichert, B. Poelsema, and G. Comsa, Phys. Rev. Lett. {\bf 71},
895 (1993).

\bibitem{chambliss}
 D. D. Chambliss and K. E. Johnson, Phys. Rev. B {\bf 50}, 5012 (1994).

\bibitem{kryukov}
Y. A. Kryukov and J. G. Amar, Phys. Rev. E {\bf 83}, 041611 (2011);
B. C. Hubartt, Y. A. Kryukov, and J. G. Amar, Phys. Rev. E {\bf 84}, 021604 (2011).


\end{references}
\end{document}